\documentclass[iop,apjl]{emulateapj}

\slugcomment{Draft v6.4}

\shorttitle{Reverberation Mapping of A Protoplanetary Disk}
\shortauthors{Meng et al. 2016}

\begin{document}
\title{Photo-reverberation Mapping of a Protoplanetary Accretion Disk\\around a T Tauri Star}

\author{Huan Y. A. Meng\altaffilmark{1,2,3}, Peter Plavchan\altaffilmark{1,4}, George H. Rieke\altaffilmark{2,3}, Ann Marie Cody\altaffilmark{5}, Tina G\"{u}th\altaffilmark{6}, John Stauffer\altaffilmark{7}, Kevin Covey\altaffilmark{8}, Sean Carey\altaffilmark{7}, David Ciardi\altaffilmark{1}, Maria C. Duran-Rojas\altaffilmark{9}, Robert A. Gutermuth\altaffilmark{10}, Mar\'{i}a Morales-Calder\'{o}n\altaffilmark{11}, Luisa M. Rebull\altaffilmark{7}, Alan M. Watson\altaffilmark{12}}
\email{hyameng@lpl.arizona.edu}

\altaffiltext{1}{Infrared Processing and Analysis Center, California Institute of Technology, MC 100-22, 770 S Wilson Ave, Pasadena, CA 91125}
\altaffiltext{2}{Lunar and Planetary Laboratory and Department of Planetary Sciences, University of Arizona, 1629 E University Blvd, Tucson, AZ 85721}
\altaffiltext{3}{Steward Observatory and Department of Astronomy, University of Arizona, 933 N Cherry Ave, Tucson, AZ 85721}
\altaffiltext{4}{Department of Physics, Astronomy and Materials Science, Missouri State University, 901 S National Ave, Springfield, MO 65897}
\altaffiltext{5}{NASA Ames Research Center, Moffett Field, CA 94035}
\altaffiltext{6}{Department of Physics, New Mexico Institute of Mining and Technology, 801 Leroy Pl, Socorro, NM 87801}
\altaffiltext{7}{Infrared Science Archive and Spitzer Science Center, Infrared Processing and Analysis Center, California Institute of Technology, MC 314-6, 1200 E California Blvd, Pasadena, CA 91125}
\altaffiltext{8}{Department of Physics and Astronomy, MS-9164, Western Washington University, 516 High St, Bellingham, WA 98225}
\altaffiltext{9}{Instituto de Astronom\'{i}a, Universidad Nacional Aut\'{o}noma de M\'{e}xico, Apartado Postal 106, 22800, Ensenada, Baja California, M\'{e}xico}
\altaffiltext{10}{Department of Astronomy, University of Massachusetts, Amherst, MA 01003}
\altaffiltext{11}{Centro de Astrobiolog\'{i}a, Departamento de Astrof\'{i}sica, INTA-CSIC, PO Box 78, E-28691, ESAC Campus, Villanueva de la Ca\~{n}ada, Madrid, Spain}
\altaffiltext{12}{Instituto de Astronom\'{i}a, Universidad Nacional Aut\'{o}noma de M\'{e}xico, Apartado Postal 70-264, 04510 M\'{e}xico, D. F., M\'{e}xico}
\altaffiltext{}{}
\altaffiltext{}{}
\altaffiltext{}{}

\begin{abstract}
Theoretical models and spectroscopic observations of newborn stars suggest that protoplantary disks have an inner ``wall'' at a distance set by the disk interaction with the star. Around T Tauri stars, the size of this disk hole is expected to be on a 0.1-AU scale that is unresolved by current adaptive optics imaging, though some model-dependent constraints have been obtained by near-infrared interferometry. Here we report the first measurement of the inner disk wall around a solar-mass young stellar object, YLW 16B in the $\rho$ Ophiuchi star-forming region, by detecting the light travel time of the variable radiation from the stellar surface to the disk. Consistent time lags were detected on two nights, when the time series in $H$ (1.6 $\micron$) and $K$ (2.2 $\micron$) bands were synchronized while the 4.5 $\micron$ emission lagged by $74.5\pm3.2$ seconds. Considering the nearly edge-on geometry of the disk, the inner rim should be 0.084 AU from the protostar on average, with an error of order 0.01 AU. This size is likely larger than the range of magnetospheric truncations, and consistent with an optically and geometrically thick disk front at the dust sublimation radius at $\sim$1500 K. The widths of the cross-correlation functions between the data in different wavebands place possible new constraints on the geometry of the disk.
\end{abstract}



\section{Introduction}

Stars form by fragmentation and gravitational collapse within giant molecular clouds \citep{dun14}. Due to the conservation of angular momentum, in the first few million years after star formation, most young stellar objects (YSOs) still host accreting protoplanetary disks of relatively primordial gas and dust, which are later dispersed to give birth to planets \citep{wil11}. The structure of the inner region of a protoplanetary disk depends on the mechanism by which the disk mass accretes onto the central star. One of the crucial diagnostics for various models is where the inner disk is truncated. The inner boundary of a protoplanetary accretion disk is expected to be shaped by sublimation \citep[for dust,][]{mon02,muz03} and/or stellar magnetospheres \citep[for gas,][]{kon91}, forming an optically thick inner disk ``wall''.

The current paradigm of the inner disk structure is largely constrained by near-infrared interferometry \citep{mil07,dul10,ant15}, from which the measurements of the inner disk rims are to some extent model-dependent \citep[e.g.,][]{ake05}. This works well for larger disk holes around higher mass Herbig Be and Ae stars. However, the inner disk holes of solar-mass T Tauri stars are expected to be on the order of 0.1 AU. This is too small to be directly resolved by current adaptive optics imaging, while it also makes the interpretation of interferometric observations sensitive to model assumptions, leading to controversial results \citep[cf.][]{eis07,pin08}.

For new insights to the disk structure, we have probed the inner disk region in a relatively model-independent way, by measuring the light travel time from the protostar to the inner wall of its disk \citep{har11,pla16}. We introduce the observations and data reduction procedures in \S2, analyze the observational results in \S3, and finally discuss the accretion mechanism and other constraints in \S4. A summary is given in \S5.

\section{Observations and Data Reduction}

We selected L1688 \citep[a part of the $\rho$ Ophiuchi star-forming region at $\sim$120 pc,][]{wil08} as the target of this experiment, because of its proximity and high concentration of variable YSOs in the 2MASS calibration field observations \citep{pla08,par14}. Ideally, we should monitor the stellar variations at visible wavelengths. However, the L1688 field is behind an interstellar cloud with high extinction \citep[$A_V \sim 30$,][]{eva03}. The shortest wavelength that can yield sufficient signal-to-noise ratio (S/N) while allowing high exposure cadence is $H$ band in the near infrared. Therefore, we used the near-infrared $H$ and $K$ bands with 4 ground-based telescopes, together with coordinated {\it Spitzer Space Telescope} observations to monitor the YSOs.

\subsection{{\it Spitzer} and Ground-based Observations}

The {\it Spitzer} observations were made in its warm mission with the Infrared Array Camera \citep[IRAC,][]{faz04} under program 60109 for three periods of staring mode observations at 4.5 $\micron$, on April 20, 22, and 24, 2010, respectively. Each session lasted about 8 hours, with frame time of 2 seconds and no dithering. We chose 4.5 $\micron$ over 3.6 $\micron$ with IRAC to maximize the fractional flux contribution from the disk.

\begin{figure}
\includegraphics[scale=.6,angle=0]{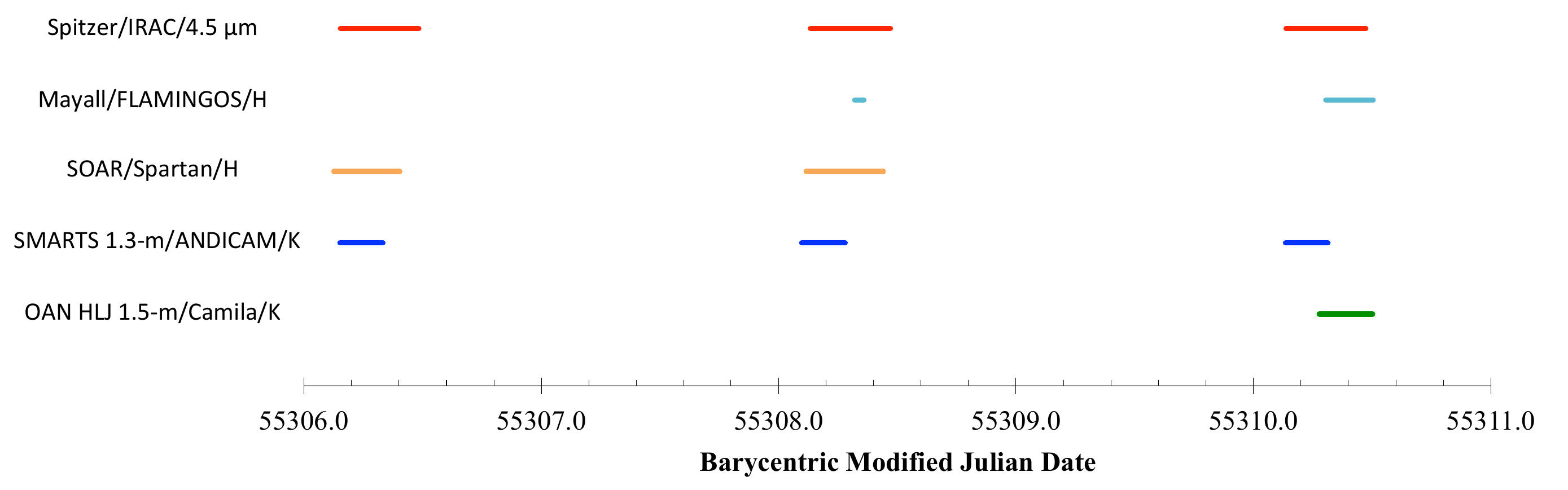}
\caption{Time coverage of each participating telescope. Intra-night gaps in coverage are not shown. The Mayall/FLAMINGOS data from the second night and the {\it Spitzer}/IRAC data from the third night are not used because of too short time coverage and over-saturation, respectively.\label{fig1}}
\end{figure}

\begin{figure}
\epsscale{1}
\plotone{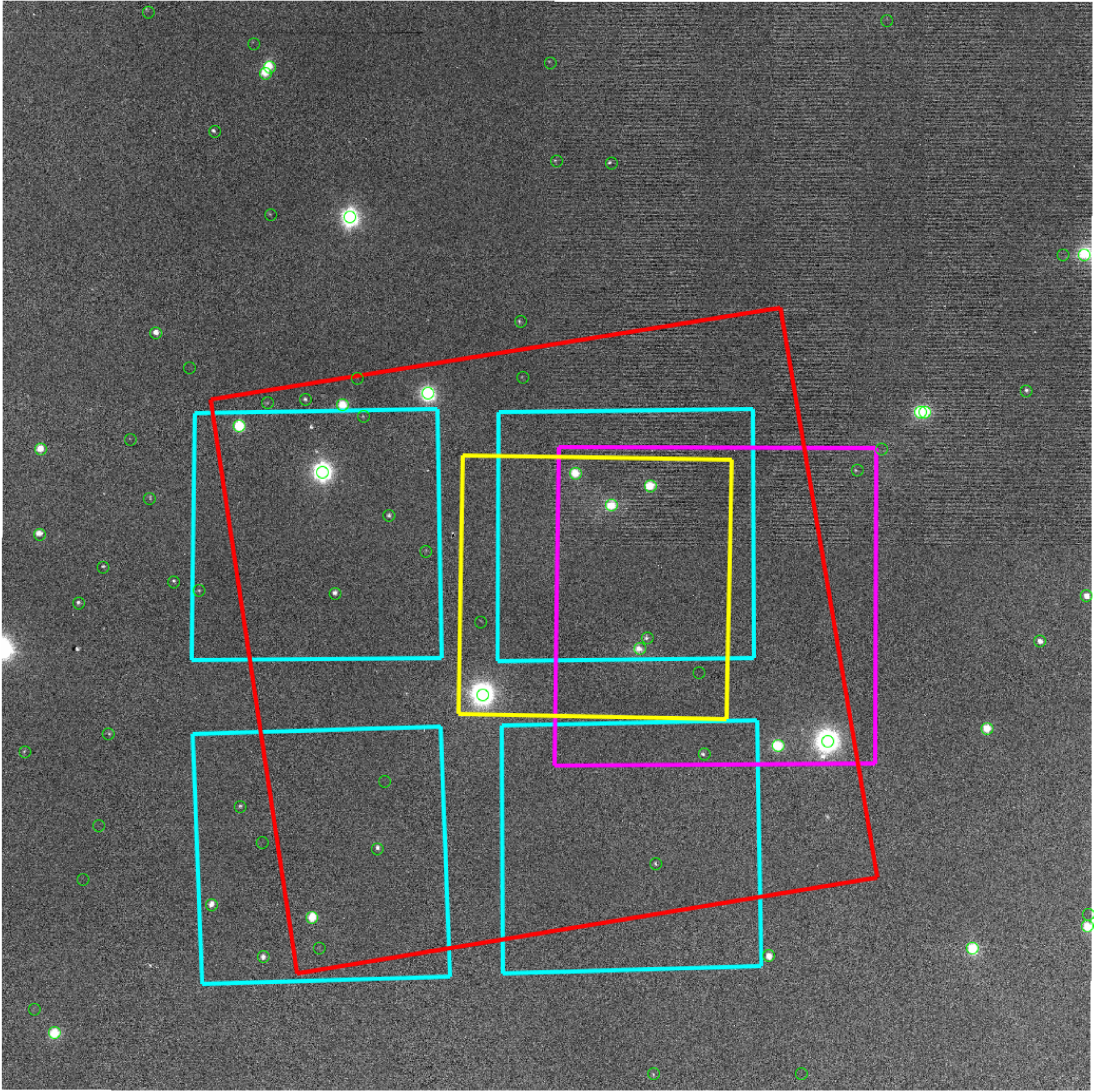}
\caption{Fields of view of the participating telescopes. The background is a full image taken by Mayall/FLAMINGOS in $H$ band, with all 2MASS sources circled in green. North is up and east is to the left. Overlapped are the fields of view of {\it Spitzer}/IRAC (red), SMARTS 1.3-m/ANDICAM (yellow), OAN Johnson 1.5-m/Camila (magenta), and SOAR/Spartan (cyan). SOAR/Spartan consisted of four separate detectors, and thus provided a discontinuous field of view.\label{fig2}}
\end{figure}

Near simultaneous ground-based observations were made with the FLAMINGOS on the 4-m Mayall telescope at Kitt Peak National Observatory, Arizona; the Spartan Infrared Camera on the 4-m SOAR telescope and the ANDICAM on the SMARTS 1.3-m telescope, both at Cerro Tololo Inter-American Observatory, Chile; and Camila on the Harold L. Johnson 1.5-m telescope of Observatorio Astron\'{o}mico Nacional (OAN) at Sierra San Pedro M\'{a}rtir, Mexico. We used the larger telescopes, Mayall/FLAMINGOS and SOAR/Spartan, in $H$ band and the smaller telescopes, SMARTS 1.3-m/ANDICAM and OAN Johnson 1.5-m/Camila, in $K$ band. The time coverage for each telescope is illustrated in Figure~\ref{fig1}. The field of view of each participating telescope, including {\it Spitzer}, is illustrated in Figure~\ref{fig2}.

Twenty-seven sources that are in the 2MASS catalog \citep{skr06} were detected in the {\it Spitzer}/IRAC field of view. All these sources were covered by ground-based observations on at least one night. For all observations, we relied on the original time in the telescope systems as recorded in the image headers. To correct the light travel time among participating telescopes at different locations, including {\it Spitzer}, the mid-point of each exposure is converted to the barycentric modified Julian date in dynamical time (BMJD\_TDB) based on the celestial coordinates of the field.

\subsection{Data Reduction}

\subsubsection{Photometry}

The ground-based data were prepared with bias, dark, and flat field corrections, where the flat fields used were ``super-sky flats'' constructed based on the science images themselves by rejecting stars and extreme pixel values. Aperture photometry was performed on each of the processed science images. To guarantee overlapped coverage with the {\it Spitzer} observations, the ground-based observations were not always made under photometric weather, and no absolute calibration was attempted. Instead, we mutually compared the time series of the instrumental magnitudes to look for isolated and unsaturated bright stars in the common field of view of all telescopes, and finally selected three sources as the reference calibrators for differential photometry: YLW 15A, YLW 16A, and YLW 16C. The selection of the same set of reference stars for all data sets is important for comparing the light curves from different telescopes at different wavelengths. Since the three stars are similar in brightness in both $H$ and $K$ bands, we adopted the unweighted average of their instrumental magnitudes as the calibration reference for all sources observed, including the three stars themselves.

For the {\it Spitzer} data, we used the automated IDL package Cluster Grinder \citep{gut09} to perform aperture photometry. This pipeline was used in other works of the YSOVAR project \citep[e.g.,][]{gun14}. However, in contrast to other YSOVAR monitoring, the {\it Spitzer} observations in this work were designed to be in staring mode for $\sim$8 hours with no dithering and no break between consecutive exposures, making it irrelevant to combine mosaics and define epochs. Hence, we only relied on the photometry on the BCD images to form the time series and retain the highest possible time resolution.

A complete data set for the reverberation experiment is provided in Table~\ref{database}, with Table~\ref{index} as the index of the stars. It includes the differential photometry for all twenty-seven sources monitored by all participating ground-based telescopes over all three nights. Accurate {\it Spitzer} photometry for the sources is subject to corrections for saturation and intra-pixel sensitivity variations, which need to be addressed case by case and is beyond the scope of the paper. For interested readers, the {\it Spitzer} data are publicly available through NASA/IPAC Infrared Science Archive under program 60109.

Although some of the sources are known infrared variables on longer timescales, most of them did not vary significantly or had variations uncorrelated at different wavelengths on a timescale less than 8 hours. Only one source, the class I \citep{gut09} \citep[or edge-on class II,][]{van09} member known as YLW 16B (= IRS 46 = ISO-Oph 145 = GY 274 = 2MASS J16272943-2439161), had significant hourly variations in all three wavebands that appear to be mutually correlated. The rest of the work is focused on this object.

\subsubsection{Corrections for IRAC Intra-pixel Sensitivity Variations}

Staring mode IRAC observations are known to be affected by intra-pixel variations in detector response. Several approaches have been introduced to correct these effects, e.g., non-parametric sensitivity mapping \citep{bal10}, polynomial fitting of stellar centroid \citep{rea05,knu08}, and power spectrum minimization \citep{cod11}, see \citet{irac2015}. None of the methods are applicable to saturated data, while YLW 16B was saturated on the BCD images on all three nights of our observations. To deal with the pixel phase effect and the saturation, we made photometry with 2 different apertures on the BCD images: a small aperture with a radius of 2.8 pixels, and a larger aperture of 5.6 pixels. Considering that an overexposed point spread function (PSF) should first saturate the pixels around the centroid, we subtracted the small aperture flux from the large counterpart to obtain the brightness of the annular PSF wing, which is less prone to saturation and is fainter than the entire stellar flux by a constant factor only dependent on the instrumental PSF. Though with reduced S/N, an important benefit of this method is that the PSF wing contains more pixels with more even illumination, and thus is much less affected by the intra-pixel variations, so we needed to make no corrections for this effect.

On the first night, the opto-center of the YSO coincidentally lay close to and wandered around the corner of 4 adjacent pixels. We found the photometry values offset and noisier when the PSF centroid was on the northwest pixel, and chose to reject the corresponding part of the data. On the second night, the photometry of the PSF wing worked well for all the data. Unfortunately, on the third night our approaches for de-trending the pixel phase effect were not successful, likely due to stronger saturation and the particular pixel location where the star fell on this night; we had to discard the 4.5 $\micron$ data. The final light curves, after corrections for the IRAC pixel phase effect were applied, are shown in Figure~\ref{fig3}.

\subsection{Cross Correlations}

As a general strategy to maximize the time resolution, the experiment design requested all participating telescopes, including {\it Spitzer}, to take new exposures immediately after the previous image was read out. The resultant mean cadence was 13 seconds for OAN Johnson 1.5-m/Camila, 17 seconds for SOAR/Spartan, 20 seconds for Mayall/FLAMINGOS and SMARTS 1.3-m/ANDICAM, and 3.4 seconds for {\it Spitzer}/IRAC. None of the instruments yielded evenly spaced time series, but the variations of the sampling frequency of each instrument were small (mostly less than 1 second with occasional omissions of a data point) and appeared to be random. These should have no effect on our science results.

To preserve all information in the original data, before computing each cross-correlation function (CCF), we used the faster of the average cadences in the two compared time series as the equidistant time step to interpolate both data sets, except for skipping the periods that one or both time series did not cover. Then, we computed the CCFs between the time series in $H$ and $K$ bands and at 4.5 $\micron$. Each CCF within $\pm$170 seconds (time for 100 {\it Spitzer} BCD frames) was fitted by a Gaussian to look for any time lag, and by a quadratic polynomial for a sanity check. The time lags estimated this way are expected to be more robust \citep{zha06}, and do not necessarily coincide with the exact location of a local minimum, which is more prone to noise influence.

For CCFs involving 4.5 $\micron$ data, we achieved the most accurate time lag measurements on the second night. Because of the lower variation amplitudes and shorter overlaps of time coverage, on the first night the peaks of the CCFs were shallower and worse defined, resulting in less accurate time lag measurements. To validate the reality of the lag, we performed the following test. At each wavelength, we first fitted the light curve with a cubic spline function and subtracted it from the data, giving two signal streams: (1) the fitted variations without noise; and (2) the time-series photometric noise de-trended for the source variations. We then added these two signals together with different artificial phase lags and extracted the CCFs from the artificial combined signal. The results suggested that the detected time lags between each pair of wavelengths have a 1$\sigma$ uncertainty of $\sim$3.9 seconds and are free from noise aliasing.

We found that the light curves of YLW 16B in $H$ and $K$ bands were consistently synchronized over all three nights with an insignificant average offset of $-1.4\pm2.4$ seconds; the 4.5 $\micron$ signals lagged behind both $H$ and $K$ by $74.5\pm3.2$ seconds, accurately measured on the second night and consistent within 2$\sigma$ errors with the less constraining data from the first night (Figure~\ref{fig4}).

\section{Variability Analysis}

The hourly photometric variations seen in YLW 16B could arise from various components in the system, such as the protostar itself or zones in the circumstellar disk. Such variations could be correlated. Increases in the stellar or accretion luminosity should, for example, heat the disk.

Fortunately, our interpretation of time lags can be simplified because the intrinsic responses should be quick. Reflection of variations from other structures should be instantaneous once allowing for light travel time. Thermal variations will also be fast. In the innermost region of a young accretion disk, dust particles are typically small \citep[on $\micron$ scales,][]{mon02,mcc13} and have little thermal capacity. Their thermal relaxation time should be well under a second, much faster than the time resolution we can achieve in real observations. Therefore, to first order we can attribute any time lag between the spectral bands to light travel time (although radiative transfer may become important on longer timescales).

Since the observed lag is expected to be the light travel time, it is critical to identify the source of the variable component and the response mechanism at each wavelength. For this purpose, we looked into two aspects of YLW 16B: the near-infrared spectrum, and the characteristics of the variations.

\subsection{Near-Infrared Spectrum}

To verify the detection of the central star and constrain its spectral type, we utilized the high-resolution spectrum taken with Keck/NIRSPEC on May 30, 2000 \citep[R$\sim$18000,][]{dop05} to look for stellar absorption features in $H$ and $K$ bands. Although \citet{dop05} did not identify photospheric lines in the NIRSPEC spectra of YLW 16B, after careful inspection of the data we felt that several such lines might exist. To test their reality, the spectrum was cross-correlated with a BT-Settl synthetic model spectrum \citep{all12} from 1800 to 6000 K in increments of 200 K with solar metallicity and gravity $\log g = 4.5$. Both observed and synthetic spectra were normalized. The NIRSPEC spectrum had higher spectral resolution than the BT-Settl template, so we boxcar smoothed it over 3 pixels, which made the data less noisy and with comparable resolution to the model. We then identified the stellar temperature and wavelength shift that yielded the maximum cross-correlation. As shown in Figure~\ref{fig5}, the largest cross-correlation appears to be at 4200 K in the spectral order around 2.1 $\micron$. However, a closer inspection found inconsistency with other wavebands in both temperature and shift. Combining all three orders, the spectrum suggests the most likely effective temperature of $3600\pm600$ K, corresponding to a spectral type of $\sim$M2. Adopting a uniform spectral calibration provides solid identifications of the \ion{Na}{1} absorption triplet at 2.21 $\micron$, and the \ion{Mg}{1} and \ion{Al}{1} absorption lines at 2.09 $\micron$ at less significance.

We experimented with line depths scaled to values from 5\% to 25\% to accommodate a corresponding range of star-to-disk flux ratios and possible emission filling-in of the absorption lines. However, we found that the cross-correlation produced no useful constraint to the line depth scale.

As a sanity check, we compared the findings with the lower-resolution spectra in $H$ and $K$ bands, observed by IRTF/SpeX weeks before and after the reverberation experiment on March 27, April 1, April 2, and May 7, 2010 \citep{fae12}. Before cross-correlation, the spectra observed at all 4 epochs were combined for higher S/N. The synthetic BT-Settl templates were smoothed to the SpeX spectral resolution (R$\sim$1200). Both data sets were normalized. The cross-correlation, as shown in Figure~\ref{fig5}, did not yield any obvious peak. However, the maximum also occurred at $T_{eff}$ = 3600 K with consistent wavelength calibrations for both $H$ and $K$ bands, though at a weak significance (cross-correlation $\rho \sim 0.25$).

In addition, the combined SpeX spectrum also reveals several atomic emission lines that are typical for evaporating disks with disk winds (Figure~\ref{fig6}). The most prominent are Br $\gamma$ of \ion{H}{1} at 2.17 $\micron$, the \ion{Mg}{1} lines at 1.49 and 1.50 $\micron$, and the \ion{Na}{1} lines at 2.21 $\micron$. Interestingly, the 2.21 $\micron$ \ion{Na}{1} lines were absorption features in the NIRSPEC spectrum, but intermittently appeared as emission lines with variable equivalent widths in the SpeX spectra. This suggests that emission filling of absorption lines plays a critical role in the detectability of some photospheric absorption features. Therefore, the line depth scale may not be a valid constraint to the fractional flux contribution in this case. The most reliable flux ratios for the star and the disk may have to be estimated by other means.

\subsection{Color-Magnitude Diagram and Variability Mechanisms}

We used a color-magnitude diagram to identify the mechanisms underlying the observed variations. Because the primary energy source of the system is the central protostar and the 4.5 $\micron$ photometry is most contaminated by the radiation of the disk, we focused on the $H$ and $K$ data, leaving us $K$ vs $H-K$ as the only option for a color-magnitude diagram. Because $K$ and $H-K$ are not independent, errors in $K$ will affect the $H-K$ color, producing a false negative slope. To reduce this effect to an insignificant level while retaining some time information, we averaged the $H$ and $K$ photometry every 204 seconds. The range is defined as 60 times the fastest average cadence in the experiment (3.4 seconds). Such averaged photometry was plotted in the color-magnitude diagram in Figure~\ref{fig7}. The nominal errors were 0.002 magnitude in both $K$ and $H-K$.

In $H$ and $K$ bands, an important reference is the 2MASS calibration field photometry of YLW 16B \citep{par14}. Over a time span of $\sim$900 days, we found that the YSO varied between 9.25 and 11.57 magnitude in $K_s$, and between 2.55 and 3.20 in $H-K_s$. Tracking the variations in the color-magnitude diagram of $K$ vs $H-K$, we found mixed behaviors of YLW 16B that can be classified into three types. The long-term variations of the YSO form a locus of positive slopes with moderate scatter over the time span of the 2MASS calibration field observations; the relative color-magnitude relation between the first two nights of our reverberation observations (for which we have comparable $H$ and $K$ photometry and good 4.5 $\micron$ data) had a negative slope, also with moderate scatter; finally, the hourly light curve for each night appears to have mixed behavior with no dominant trend.

Such complexity indicates that different variability mechanisms are at work on different timescales. We considered a number of possible variability mechanisms. (1) Variable extinction. Both interstellar and circumstellar extinction may vary with time \citep{car01,cod14}. The interstellar extinction law has been well measured towards the Galactic plane and several star-forming regions \citep{ind05,fla07,cha09}. Thanks to ongoing dust growth, circumstellar dust around YSOs may have a different size distribution and optical properties from its interstellar counterpart \citep{mcc13}, possibly leading to a different extinction law. However, because the Keplerian timescale of the inner disk is orders of magnitude shorter than the time needed for global disk evolution, circumstellar extinction is expected to vary quasi-periodically \citep{hil13,sta15}. (2) Starspots on the stellar photosphere \citep{gue93,ngu09,poz15}. Magnetic activity within the YSO can trigger cool starspots like sunspots, and the accretion mass flow from the disk through the magnetic field lines may produce accretion ``hotspots'' \citep{sta14} near the magnetic poles on the photosphere of the protostar \citep{har94}. Variability associated with starspots should be on the timescale of the rotation period of the protostar. Radiation from starspots is usually considered as blackbody, but with different temperatures from the rest of the photosphere. (3) The accreting gas along the magnetic field lines just above the photosphere of the protostar. Such material is supposed to be the precursor of accretion hotspots. Differences in spectral properties are that the accreting gas can be optically thin on its way to the star (free-free emission), and does not have to occupy or obscure any area of the stellar photosphere because of the edge-on line-of-sight, even if it is optically thick (free-bound or blackbody-like emission). (4) Variations of the observable thermally emitting area of the inner disk wall. The temperature of the inner disk wall is likely close to the dust sublimation temperature \citep{mon02,muz03} and radiates like a blackbody. The timescale for the inner disk variability should be quasi-Keplerian.

We plotted and compared all the observations and theoretical models in the color-magnitude diagram in Figure~\ref{fig7}(A). Each data point from the reverberation observations, averaged over 204 seconds,  is individually plotted. The errors are smaller than the symbol sizes and are not drawn. Although these observations were not calibrated to the 2MASS standard, they are self-consistent and are valid for relative comparisons. The interstellar extinction vector is labeled with the 2MASS extinction law \citep{fla07}. The 2MASS data are divided into 5 bins of $K_s$ magnitude, from 9.0 to 11.5 with a bin size of 0.5. The median $H-K_s$ color of each bin is plotted, with the first and third quartiles indicated by horizontal bars.

The exact placement of the spotless protostar on the color-magnitude diagram is poorly constrained and subject to the uncertainties in extinction. In the diagrams, we place it arbitrarily for easier comparisons. The stellar photosphere is considered as a 3600 K blackbody. The solid lines of different colors represent the models of the variability mechanisms considered, as labeled in the legend. The cool and hot starspots are assumed to be 2000 and 8000 K, respectively. Three temperatures for the optically thick (blackbody-like) accreting gas are plotted, 6000, 10000, and 14000 K. For the cool and hot starspots and the optically thick accreting gas models, each line is drawn up to the point when the emitting areas of the added components are 40\% of the area of a hemisphere of the stellar photosphere, no matter whether the photosphere is occupied (starspots) or not (accreting gas). The model line for variable dust emitting area is plotted to the point at which the disk emitting area is 12 times larger than a hemisphere of the stellar photosphere.

From Figure~\ref{fig7}, it is clear that the overall positive slope (note the inverted Y-axis) of the long-term variations in the 2MASS data primarily reflects the variations of activities related to the protostar, like starspots or optically thick accreting gas, since those are the only models that have potentially matching positive slopes. The scatter in 2MASS $H-K_s$ is similar in amplitude to the amplitude of the short-term variations in the reverberation experiment, implying that changes similar to those seen in our monitoring have occurred frequently in the past.

The negative slope, as seen between the first two days of the reverberation experiment, appears to be caused by variation in the observable disk emitting area. To verify this hypothesis, we referred to the time-resolved information of the IRTF/SpeX spectra at the 4 quasi-contemporary epochs. In the flux calibrated data, shown in Figure~\ref{fig8}(A) and (B), we noticed that the spectra on March 27 and May 7 had practically identical fluxes and spectral slopes in $H$ band, with synthetic photometry agreeing to within 2\%, probably better than the uncertainties. But the March 27 data were significantly brighter and redder in $K$, providing the basis to compare two different states of the inner disk. We subtracted the $K$ band spectrum on May 7 from that on March 27, and de-reddened the difference with $A_V = 30$. We found that the differential spectrum, as shown in Figure~\ref{fig8}(C), does not agree with optically thin free-free emission, but can be reasonably fitted with a blackbody. (The best-fit temperature is 1100 K, but this is only a very rough estimate because of the calibration uncertainties and limited wavelength coverage from which it is derived.) This reveals the spectrum ``added'' to K band on March 27 that made it brighter and redder than on May 7, and confirms that a variable disk emitting area is likely responsible for the negative slope in Figure~\ref{fig7}.

After associating the long-term and daily variations to their respective model components, now we focus on the hourly variations that led to the detected time lag. A zoom-in view of the color-magnitude diagram of the reverberation data is shown in Figure~\ref{fig7}(B). Despite a significant day-to-day shift, the intra-day observations appear to have mixed behavior with no dominant color trend. The squared linear correlation coefficients are $\sim$0.001 for the data of both nights.

The evolution between consecutive data points (separated by 204 seconds) is reversible back and forth in the diagram. The disk emitting area could not possibly increase and then decrease on such short timescales, making the model incompatible with the hourly changes. Variable extinction may play a role over any timescale. But the fixed extinction vector alone cannot explain the random directions in the color-magnitude evolution. The most likely interpretation for the hourly variations is that they are caused by a changing continuum from the accreting gas \citep{gue93,sta14}. The net effect on the output of the source is a varying combination of optically thick and optically thin emission, or that the optical depth was varying in an intermediate range. Variable extinction may still play a role, but even if it did, it does not introduce any time lag between different wavelengths. Therefore, the source of the variable part of the YLW 16B radiation should geometrically be in the magnetically confined accreting columns right above the photosphere of the protostar.

Given the edge-on geometry of YLW 16B, the disk emission we see should mostly come from the far side of the inner rim \citep{lah06,bas13}. Therefore, to have the direct stellar light curve (from the near side) and responding disk light curve (from the far side) correlated, the source of the variability signals must be seen from both sides. The only satisfactory place for this source is near the limb on the star. Such visibility can be naturally explained if the variability signals are from the accreting gas, because the accretion funnels should hit the star near the magnetic poles, which are likely near or even aligned with the stellar rotation axis. These areas should always be near the limb of the edge-on star as seen from the observer.

\subsection{Relative Flux Contribution of Each Source}

Building a spectral energy distribution (SED) model that fits the available measurements of YLW 16B at all time is beyond the scope of this paper. Instead, with this analysis we aim to dissect the YSO to estimate the flux contribution of each emitting entity at our working wavelengths at the epoch of the reverberation experiment. This analysis will help us understand the observations of variations among the source components.

\subsubsection{Color Blending}

Considering the highly variable nature of the source, to deconvolve the colors of the individual source components we should only adopt the data taken at the time of the reverberation experiment. However, using our photometry ignores the color terms for the filters and for the source SEDs, which are not possible to be corrected given the strong variability of the source. Therefore, the $H$ and $K$ band photometry of the reverberation observations is not precisely on the 2MASS standard scale. The differences in bandpass may introduce systematic errors, because the extinction law and colors of young stars are established based on the 2MASS definitions of $JHK_s$ wavebands. To overcome the problem, we chose to use the 2MASS photometry, but only picked values that are representative of the state of the YLW 16B system at the time of the reverberation experiment. 

For this purpose, we compared the averaged reverberation $H$ and $K$ magnitudes to their counterparts in the multi-epoch 2MASS calibration field photometry \citep{par14} by computing their combined magnitude differences, defined as
\begin{equation}
M_D = \sqrt{(H_{reverb} - H_{2MASS})^2 + (K_{reverb} - K_{s,2MASS})^2}
\end{equation}
The average magnitudes of the reverberation observations are $H = 12.14$, $K = 9.29$, and $[4.5] = 5.75$. The smallest difference in all 2MASS entries is $M_{D,min}$ = 0.132 magnitude, with the color of
\begin{eqnarray*}
(J-H)^{obs} & = & 4.122 \pm 0.046 \\
(H-K_s)^{obs} & = & 2.663 \pm 0.014
\end{eqnarray*}
There are 21 sets of 2MASS measurements with $M_D \leq 3 M_{D,min}$. Averaging all of them yields essentially the same color indices within errors, which means the color is insensitive to our selections. In the following calculations we used the single 2MASS measurement with the minimum $M_D$, and assumed that it resembles what the reverberation photometry should be if calibrated to the 2MASS standard. In addition to reconciling the definitions of the wavebands, another important benefit of using the 2MASS data is the inclusion of $J$ band photometry, which is more sensitive to the extinction and stellar output and thus provides better constraints on the component deconvolution.

The observed color of the YSO is made up of three components, the radiation from the stellar photosphere, from the inner disk, and from the accreting gas, all behind the foreground extinction along the line of sight, yielding
\begin{eqnarray}
\left( \frac{F_J}{F_H} \right)^{obs} & = & \frac{10^{-0.4 A_J} \left( F_J^* + F_J^{disk} + F_J^{gas} \right)}{10^{-0.4 A_H} \left( F_H^* + F_H^{disk} + F_H^{gas} \right)} \label{JH_obs} \\
\left( \frac{F_H}{F_{K_s}} \right)^{obs} & = & \frac{10^{-0.4 A_H} \left( F_H^* + F_H^{disk} + F_H^{gas} \right)}{10^{-0.4 A_{K_s}} \left( F_{K_s}^* + F_{K_s}^{disk} + F_{K_s}^{gas} \right)} \label{HK_obs}
\end{eqnarray}
where $F$ is flux density, and $A$ is the extinction in magnitudes. Subscripts and superscripts denote the wavebands and sources of radiation, respectively. Although the adopted set of 2MASS photometry is the closest in brightness to our reverberation observations, to be conservative, in the equations we did not connect it to the 4.5 $\micron$ reverberation photometry because of the non-simultaneity. The averaged 4.5 $\micron$ photometry of the reverberation experiment is left for a consistency check of the $JHK_s$-based models.

We took the colors of young stellar photospheres from observations \citep{luh10}. Given M2 as the approximate spectral type of YLW 16B, we adopted the intrinsic color of the photosphere as
\begin{eqnarray*}
(J-H)^* & = & 0.70 \\
(H-{K_s})^* & = & 0.20 \\
({K_s}-[4.5])^* & = & 0.17
\end{eqnarray*}
The radiation of the inner disk should be close to a blackbody \citep{mon02,muz03}. Assuming a temperature of 1100 K (see \S3.2 for the temperature estimate), the disk color should be
\begin{eqnarray*}
(J-H)^{disk} & = & 2.47 \\
(H-{K_s})^{disk} & = & 1.58 \\
({K_s}-[4.5])^{disk} & = & 2.51
\end{eqnarray*}
We find that the extinction law specifically determined in $\rho$ Ophiuchi is not accurate enough for our purpose \citep{eli78,ken98}. Fortunately, the infrared extinction law of $\rho$ Ophiuchi is practically identical to that of other star-forming regions \citep{har78,fla07}. Here we adopted the generic extinction law for 2MASS and {\it Spitzer} wavebands \citep{fla07,cha09}, where
\begin{eqnarray}
A_J & = & 2.5 A_{K_s} \\
A_H & = & 1.55 A_{K_s} \\
A_{[4.5]} & = & 0.53 A_{K_s}
\end{eqnarray}
We assume $S$ as the disk emitting area in units of the area of the observable surface of the protostar. Then, we can construct all fluxes by introducing proper ratios of the fluxes between different wavebands, sources, and temperatures. If a flux ratio is $\eta_{V_1/V_2}^C$, where $V_1$ and $V_2$ are the varied conditions (e.g., wavelengths or temperatures) and $C$ is the common condition, the unreddened flux at the $i$-th waveband in a set of $n$-band photometry can be written as
\begin{eqnarray}
F_{\lambda_i}^* & = & E_{\lambda_n}^{disk} \eta_{T^*/T^{disk}}^{\lambda_n} \eta_{\lambda_i/\lambda_n}^* \\
F_{\lambda_i}^{disk} & = & E_{\lambda_n}^{disk} S \eta_{\lambda_i/\lambda_n}^{disk} \\
F_{\lambda_i}^{gas} & = & F_{\lambda_n}^* \eta_{gas/*}^{\lambda_n} \eta_{\lambda_i/\lambda_n}^{gas}
\end{eqnarray}
where $E_{\lambda_n}^{disk}$ is the disk flux emitted per unit area at the longest ($n$-th) wavelength. If the accreting gas is optically thin, its free-free emission should have a spectrum with
\begin{equation}
\eta_{\lambda_i/\lambda_n}^{gas} = \left( \frac{\lambda_n}{\lambda_i} \right)^{-0.1}
\end{equation}

Then, equation~\ref{JH_obs} and~\ref{HK_obs} become
\begin{eqnarray}
\left( \frac{F_J}{F_H} \right)^{obs} & = & \frac{10^{-A_{K_s}}}{10^{-0.62 A_{K_s}}} \left[ \frac{\eta_{J/{K_s}}^* \eta_{3600K/1100K}^{K_s} + \eta_{J/{K_s}}^{disk} S}{\eta_{H/{K_s}}^* \eta_{3600K/1100K}^{K_s} + \eta_{H/{K_s}}^{disk} S} \right. \nonumber \\
 & & \left. \times \frac{ + \eta_{3600K/1100K}^{K_s} \eta_{gas/*}^{K_s} \left( \frac{\lambda_{K_s}}{\lambda_J} \right)^{-0.1}}{ + \eta_{3600K/1100K}^{K_s} \eta_{gas/*}^{K_s} \left( \frac{\lambda_{K_s}}{\lambda_H} \right)^{-0.1}} \right] \label{JH_obs2} \\
\left( \frac{F_H}{F_{K_s}} \right)^{obs} & = & \frac{10^{-0.62 A_{K_s}}}{10^{-0.4 A_{K_s}}} \left[ \frac{\eta_{H/{K_s}}^* \eta_{3600K/1100K}^{K_s} + \eta_{H/{K_s}}^{disk} S}{\eta_{3600K/1100K}^{K_s} + S} \right. \nonumber \\
 & & \left. \times \frac{ + \eta_{3600K/1100K}^{K_s} \eta_{gas/*}^{K_s} ( \frac{\lambda_{K_s}}{\lambda_H} )^{-0.1}}{ + \eta_{3600K/1100K}^{K_s} \eta_{gas/*}^{K_s}} \right] \label{HK_obs2}
\end{eqnarray}
In equation~\ref{JH_obs2} and \ref{HK_obs2}, the flux ratios on the left are directly known from the observed color of the YSO; all $\eta$ values on the right hand side, except for $\eta_{gas/*}^{K_s}$, are known either from Planck's law or from the observation-based intrinsic color of the young stellar photosphere \citep{luh10}. Therefore, the equations are model-independent. Now we have two equations and three free parameters, $A_{K_s}$, $S$, and $\eta_{gas/*}^{K_s}$, the $K_s$-band scale factor of the emission of the accreting gas compared to the stellar photospheric flux. For any given value of $\eta_{gas/*}^{K_s}$, we can numerically solve for $A_{K_s}$ and $S$.

Taking in the 2MASS photometry with the minimum $M_D$ from the reverberation observations, we found that $A_{K_s}$ and $S$ are not sensitive to $\eta_{gas/*}^{K_s}$ within a generous range. Although the accurate value of $\eta_{gas/*}^{K_s}$ at the time of the reverberation experiment is unknown, a lower limit is constrained by the hourly peak-to-peak amplitude of 20\% observed in both $H$ and $K$ bands. At 4.5 $\micron$, thanks to the dominance of the disk emission, the same variable component may only cause a peak-to-peak amplitude $\lesssim$10\%, which would have been submerged in the noise (RMS $\sim$ 8\%) of the {\it Spitzer}/IRAC data. That being said, the amplitude of the disk response is much greater than that of the original signal. This would be naturally explained if the triggering signals are causing temperature (or chemical) changes in the disk \citep[cf.][]{fla14}.

Accordingly, for $\eta_{gas/*}^{K_s} = 0.25$, which would mean that the accretion excess can just account for the observed amplitude in $K$ band, we obtained $A_{K_s} = 3.32$ and $S = 142$; for $\eta_{gas/*}^{K_s} = 1$, which would mean that the accretion excess is as strong as that of the stellar photospheric flux in $K$ band, $A_{K_s} = 3.46$ and $S = 141$. This degeneracy is because the flux density of the stellar photosphere of a young M2 star peaks in $H$ band \citep{luh10}, causing a relatively flat SED over the $JHK_s$ wavebands, similar to that of the optically thin free-free emission assumed for the accreting gas. Extrapolating the color-blending equations to 4.5 $\micron$, for $\eta_{gas/*}^{K_s}$ values between 0.25 and 1, the predicted magnitude is in the range of 5.76 to 5.78, in good agreement with the averaged reverberation photometry $[4.5] = 5.75$. Larger values of $\eta_{gas/*}^{K_s}$ would mean that the accretion excess dominates over the stellar photospheric emission in $K$ band, and would predict fainter [4.5] magnitudes, which are not supported by the observations.

\subsubsection{Implications for Disk Properties}

Two byproducts of the color-blending equations are the extinction $A_{K_s}$ and the disk emitting area $S$. The extinction law with $R_V = 3.1$ gives $A_V/A_{K_s} = 8.8$ \citep{car89,ind05}\footnote{The extinction ratio for $\rho$ Ophiuchi is less well determined but indicates $A_V/A_K \sim 10.5$ \citep{har78,vrb93}, indicating a slightly higher extinction of $A_V \sim 35$.}. Accordingly, the $A_{K_s}$ obtained from the equations would mean $A_V \sim 30$, consistent with the extinction level of the region \citep{eva03}.

$S$ is defined as the ratio between the disk emitting area, presumably the observable surface of the inner rim, and the observable area of the protostar. Without extinction, the observable area of the protostar should be equal to its cross section. However, disk veiling over the protostar, which is not considered in the equations, may reduce the equivalent of its ``observable area'', leading to larger values of $S$. Such behavior is more likely for edge-on systems like YLW 16B. Therefore, if we know the radii of the protostar and of the inner disk hole, the value of $S$ can be used to place an upper limit on the observable area of the disk inner rim. The radius of YLW 16B was estimated to be $R_* = 2.4$ R$_{\sun}$ based on the BT-Settl model for a 1 Myr old M2 star with solar-like metallicity. Considering the contraction of the protostar as it evolves towards zero age main sequence, the estimate is in good agreement with the observed sizes of slightly older pre-main sequence stars (5-30 Myr old) with similar mass \citep{pec13}. Given the radius of the inner disk rim at 0.084 AU (as derived in the following subsection), the $S$ values obtained from the equations correspond to an upper limit of $\sim$0.2 AU, i.e., $< \pm$0.1 AU from the mid-plane, for the height of the inner disk rim if we only see the far side of it. This generous limit does not provide a useful constraint, as it can accommodate models that involve a ``puffed-up'' inner disk rim \citep{dul10}, as well as the classical, geometrically thin disk model \citep{hil92}.

\subsection{Photo-reverberation Measurement}

In summary, the variable signals in $H$ and $K$ bands should be mostly from the emission of the accreting gas near the stellar photosphere, whereas the signals observed at 4.5 $\micron$ should be dominated by the disk response. This explains the observed simultaneity between the time series in $H$ and $K$ as well as the lag at 4.5 $\micron$.

Analysis of molecular spectral features suggests that YLW 16B is close to edge-on \citep{lah06,bas13}, in which case the observed disk continuum is mostly from the far side of the inner disk wall \citep{lah06}. Such a viewing angle requires the triggering signals from the protostar to travel an additional distance before reaching us. To first order approximation, assuming the inner disk radius is azimuthally uniform, the average additional travel distance of the lagged signals should be
\begin{equation}
\frac{\displaystyle\int^{\pi}_{0} R_{in} \sin^2 \theta d \theta}{\displaystyle\int^{\pi}_0 \sin \theta d \theta} = \frac{\pi}{4} R_{in}
\end{equation}
where $\theta$ is the azimuthal angle and $R_{in}$ is the radius of the inner disk rim, which, given the observed time lag between $H$/$K$ and 4.5 $\micron$ in YLW 16B, should be $0.084\pm0.004$ AU. The nominal uncertainty of 0.004 AU only accounts for the propagated error from CCF centering and noise evalution, and should be taken as a lower limit. In converting the time lag to the inner disk radius, we had to adopt several assumptions and simplifications, such as an isotropic radiation field of the protostar, azimuthal uniformity of the disk, negligible the disk rim curvature, simplified geometry etc, which may be significant error sources beyond the observational noise. However, these errors are very difficult, if at all possible, to be quantitatively estimated with the available data. The total error of the radius of the inner disk rim is likely higher by a factor of several, on the order of 0.01 AU.

\section{Discussion}

\subsection{Disk Truncation Mechanisms}

The photo-reverberation measurement provides an independent verification of the current paradigm of the geometry and physics of the inner disk, which has been largely based on model-dependent interferometric results. Therefore, we will compare it with the expected radii from various truncation mechanisms.

To carry out this test, first we estimated the properties of the protostar and the accretion disk of YLW 16B at the time of the reverberation experiment. The total luminosity of a YSO is comprised of the intrinsic stellar luminosity and the accretion luminosity, i.e., $L_{tot} = L_* + L_{acc}$. The Br $\gamma$ emission  is a well-established indicator of accretion \citep{muz98}. We found that the equivalent width of this line is highly variable in the quasi-contemporary SpeX spectra (Figure~\ref{fig6}(A)), ranging from -2.0 \AA\ on March 27 to -4.8 \AA\ on April 2. After converting the range to Br $\gamma$ luminosity ($L_{Br \gamma}$) and compensating for extinction of $A_{K_s} = 3.3$, we used the relation \citep{nat06}
\begin{equation}
\log \left( L_{acc}/\textrm{L}_{\sun} \right) = 0.9 \left[ \log \left( L_{Br \gamma}/\textrm{L}_{\sun} \right) + 4 \right] - 0.7
\end{equation}
to derive an accretion luminosity of YLW 16B between 0.31 and 0.63 L$_{\sun}$.

On the other hand, the stellar luminosity of YLW 16B is highly uncertain. The isochrones between stellar models are significantly divergent towards younger ages and lower masses \citep{bar02,hil08}. Together with the large error in the effective temperature, we could only estimate a vast range from 0.1 to 2 L$_\sun$. To obtain better constraints, we took an alternative approach by integrating the 2MASS photometry that is best matched with the reverberation averages with the minimum $M_D$. We found that the combined 2MASS $JHK_s$ luminosity of the YSO is 0.56 L$_\sun$. Referring to the color-blending equations in \S3.2, the viable range of $\eta_{gas/*}^{K_s}$ from 0.25 to 1 means the protostar, if integrated over the $JHK_s$ wavebands, should contribute between 38\% and 51\% of the total flux. Meanwhile, we extrapolated the observed colors of young stellar photospheres \citep{luh10} with the Wien approximation (towards shorter wavelengths) and Rayleigh-Jeans law (towards longer wavelengths). For K6 to M6 stars that correspond to the range of possible temperature of YLW 16B, we found the combined $JHK_s$ flux makes 21\% to 22\% of the entire stellar luminosity (integrated from 0.1 to 5000 $\micron$), and is insensitive to the spectral type. Therefore, the stellar luminosity of YLW 16B can be narrowed down to 0.96 to 1.35 L$_\sun$. Added to the range of accretion luminosity, this makes the total luminosity of the YSO between 1.27 and 1.98 L$_\sun$. Finally, we adopted $M_* = 0.7$ M$_\sun$ for the stellar mass based on BT-Settl models, with the uncertainties allowing values from 0.5 to 1 M$_\sun$.

The accretion rate can be obtained by
\begin{equation}
\dot{M} = \frac{L_{acc} R_*}{G M_* (1 - R_*/R_{in} )}
\end{equation}
where $R_{in}$ = 0.084 AU. For the range of $L_{acc}$ from the quasi-contemporary SpeX observations, the accretion rate of YLW 16B varies between $5.4 \times 10^{-8}$ to $1.1 \times 10^{-7}$ M$_{\sun}$ year$^{-1}$ around the time of the reverberation experiment, with an average of $\sim8 \times 10^{-8}$ M$_{\sun}$ year$^{-1}$.

The magnetic field of the protostar plays a critical role in the accretion process. Though the detailed topology of the magnetic field may be complex \citep{joh14}, an important and model-independent metric is the corotation radius, $R_{cor}$, at which the Keplerian orbital period of circumstellar material matches the rotation period of the star. The corotation radius is at the balance of star-disk interaction; for accretion onto the star to proceed, the magnetospheric truncation radius has to stay interior to $R_{cor}$ \citep{bou07}. The corotation radius is given by
\begin{equation}
\frac{R_{cor}}{\textrm{1 AU}} = 1.957 \times 10^{-2} \left( \frac{M_*}{\textrm{M}_{\sun}} \right)^{1/3} \left( \frac{P_*}{\textrm{1 day}} \right)^{2/3}
\end{equation}
where $P_*$ is the stellar rotation period, which is unknown for YLW 16B. Nevertheless, given the color of $[3.6] - [8.0] = 2.42$ \citep{gut09} and the correlation between rotation periods and IRAC excessess \citep{reb06}, one stellar rotation of YLW 16B likely lasts between 6 and 10 days. Thus, the corotation radius should be in the range from 0.06 to 0.08 AU. This range, if considered as the upper limit for the magnetospheric truncation radii, is marginally compatible with our photo-reverberation measurement. Therefore, magnetospheric truncations, which should occur at even smaller radii, are unlikely to be the determining mechanism for the size of the inner disk rim of YLW 16B.

Since the dust sublimation radius depends on the total luminosity of the central star, different disk models can be compared with the observations in a size-luminosity diagram \citep{mil07}, as shown in Figure~\ref{fig9}. Based on interferometric measurements, the size-luminosity diagram has been used to suggest that some accretion disks around high mass Herbig Be stars  are ``classical''\footnote{Out of range in Figure~\ref{fig9}. See \citet{mil07}.}, i.e., optically thick but geometrically thin. In comparison,  disks around intermediate mass Herbig Ae stars can be well described by a directly heated and back-warmed ``puffed-up'' inner rim at 1500-2000 K \citep{eis04}, resulting in a weak trend toward relatively increasing apparent sublimination temperatures. However, the trend does not necessarily extrapolate to the regime of solar-mass T Tauri stars, many of which appear to have larger inner disk cavities. Oversized inner disk holes around T Tauri stars have raised questions about the roles of possibly unrecognized physical processes, such as lower dust sublimation temperatures, magnetospheric pressure \citep{eis07}, or viscosity heating \citep{muz04}.

As an independent confirmation, the reverberation inner radius of YLW 16B is larger than the model prediction for a geometrically thin ``classical'' disk, but is consistent with a ``puffed-up'' inner disk rim truncated at $\sim$1500 K in the presence of backwarming, a natural outcome of the bulk disk behind the inner wall. This resembles the disk sizes around higher mass Herbig Ae stars \citep{mil07,dul10} and does not require any additional mechanism. The result is qualitatively consistent with recent work that shows the interferometric radius of T Tauri star disks consistent with dust sublimation radii if scattered light is considered \citep{ant15}.

\subsection{Shape Constraints}\label{shape}

\citet{har11} built a radiative transfer model based on a flared disk geometry and computed the expected time lags over a range of wavelengths. By coincidence, the starting point was a hypothetical star with properties similar to those of YLW 16B: $T_{eff} \sim 4000$ K, $L_* \sim 1$ L$_\sun$, and inclination $i = 60^{\circ}$. In his model, the radius of the inner disk rim is assumed as 0.093 AU. The response signal at 4.5 $\micron$ is a combination of thermal and scattered emission from a range of stellocentric radii in the disk. The model predicts a time lag at 4.5 $\micron$ of $\sim$180 seconds, about twice the round-trip light travel time between the protostar and the disk inner rim. If this relation holds for YLW 16B, given the possible range of the total luminosity, the disk inner rim would have a radius of 0.037 AU with temperatures of 2200 - 2400 K \citep{dul10}, probably inside the corotation radius and much hotter than the typical dust sublimation temperature of $\sim$1500 K. If this were the case, the dust in the inner disk would either have to be highly refractory, or short-lived before sublimation. The disk would be truncated by the magnetosphere very close to the protostar and flared at all stellocentric distances as assumed in the \citet{har11} model. Otherwise, the disk may be self-shadowed with only a limited radial span exposed to stellar insolation, in which case the \citet{har11} model is not applicable, and the observed time lag at 4.5 $\micron$ simply corresponds to the light travel time to the effective radius of the inner disk rim.

To identify the disk geometry of YLW 16B, we noticed that the SED of the system suggests some degree of self-shadowing \citep[Figure~\ref{fig8}(D), also see the detailed modeling in][]{lah06}. Here we additionally test if the widths of the CCFs can be used to constrain the radial extent of the inner disk \citep[cf.][]{ise05,tan07}. A considerable radial span of the inner disk that is exposed to the starlight should lead to a superposition of various time lags, and thus broaden the widths of the CCFs.

A problem specific to our case of YLW 16B is that each pair of wavebands has different overlaps of time coverage, while the lengths of the overlaps and the waveform of the light curve therein could be a determining factor in yielding different widths of the CCFs. Therefore, we carried out the test by computing the CCFs based on the second night's data between 12131.0 and 19179.2 seconds in BMJD\_TDB, where we have the longest uninterrupted time coverage overlap in all wavebands (Figure~\ref{fig3}). The standard deviations of the Gaussian fits of the CCFs over this time are fairly close: 520 seconds for $H \star K$, 540 seconds for $H \star [4.5]$, and 510 seconds for $K \star [4.5]$. The similarity of all three CCF widths suggest the lack of a significant ``broadening effect'' at 4.5 $\micron$ as would be expected from a fully flaring disk \citep{dul04,har11}, and instead favors the geometry of a self-shadowed disk.

For a more quantitative evaluation, we created a set of artificially smoothed light curves in $K$ by computing the unweighted moving average of the original $K$-band time series over a window of varying length of time, and computed their CCFs with the original $H$-band data. The purpose of using both $H$ and $K$ is to mimic a time series with superposed time lags and independent noise, and assess the sensitivity of the CCF width to the degree of superposition, because the variations in the original $H$ and $K$ time series are both expected to be from the accreting gas with no time lag.

As a result, given the limited waveform and common time coverage, we found the CCF width fairly insensitive to time lag superposition. If rounded to the nearest 10 seconds, the standard deviation of the $H \star smoothed\_K$ CCF does not reach 530 seconds (the next possible value broader than that of the original $H \star K$) until $K$ is averaged over a window $>400$ seconds. To make the width of $H \star smoothed\_K$ greater than 580 seconds (i.e., 3$\sigma$ broader than those of the original CCFs), the smoothing window has to be $>950$ seconds, which is the upper limit of any ``broadening effect'' at 4.5 $\micron$. If the thermal reprocessing time of the disk is negligible, the size of the smoothing window translates to a 3$\sigma$ upper limit of $\sim$1 AU for the starlight-exposed radial extent of the disk inner edge. This limit is much smaller than the extent of the whole disk as indicated by the very red WISE color $W3 - W4 = 2.83$, corroborating a self-shadowed geometry over intermediate disk radii. However, this does not eliminate the possibility of optically thin gas and/or small amounts of refractory dust inside the rim \citep[e.g.,][]{tan08,ben10,fis11}.

As an explanation for larger interferometric disk hole sizes, it has been suggested that scattered light may inflate the interferometric measurements of the sizes of the inner disk cavities around T Tauri stars \citep{pin08}. However, the reverberation results for YLW 16B are incompatible with such scattered light models, which rely on the assumption of a flaring disk. The models require a large radial extent of the disk to contribute to the scattering \citep{pin08}, fundamentally incompatible to a self-shadowed disk geometry. Such diversity in disk geometry might also explain why the ``composite disk model'', which considers both thermal and scattered emission, provides better fits to the interferometric data of some T Tauri stars, but is ineffective in other cases \citep{ant15}.

\section{Conclusions}

To test mapping the inner disk sizes of YSOs with photo-reverberation measurements, we selected a region in L1688 in the $\rho$ Ophiuchi star-forming region, based on its proximity and high concentration of known variable YSOs. Near-simultaneous time-series photometric observations were conducted on April 20, 22, and 24, 2010, with four ground-based telescopes operating in $H$ and $K$ bands and the {\it Spitzer Space Telescope} observing at 4.5 $\micron$. Each session of {\it Spitzer} staring mode monitoring lasted $\sim$8 hours. One (YLW 16B) out of twenty-seven sources detected was found to have mutually correlated hourly variations in all three wavebands. Over all three nights, the time series measurements of YLW 16B in $H$ and $K$ bands are consistently synchronized, while the light curve at 4.5 $\micron$ lags behind both $H$ and $K$ by $74.5\pm3.2$ seconds over the first two nights when we have usable 4.5 $\micron$ data.

Analysis of the variability of YLW 16B suggests that the hourly variations likely originate from the accreting gas right above the stellar photosphere. The variable component of the flux in $H$ and $K$ bands is primarily from the accreting gas; the variability at 4.5 $\micron$ is dominated by the response of the inner disk rim. Considering the nearly edge-on, self-shadowed geometry of the disk, we derived the effective distance between the star and the inner disk rim to be $0.084$ AU on average, with an uncertainty of order 0.01 AU. The size of the inner disk rim is likely larger than required by magnetospheric accretion, but is consistent with a ``puffed-up'' disk terminated at the dust sublimation radius of $\sim$1500 K in the presence of disk backwarming. The disk truncation mechanism of YLW 16B is similar to that of higher mass Herbig Ae stars. The measurement does not agree with the model-dependent interferometric results that suggest larger than expected inner disk cavities around T Tauri stars \citep{eis07}, but is consistent with some of the recent measurements that consider scattered light \citep{ant15} The widths of the cross-correlation functions between the data in different wavebands place a 3$\sigma$ upper limit of $\sim$1 AU for the radial extent of the inner disk front.

We found that photo-reverberation is a viable technique to explore unresolved protoplanetary accretion disks. The method can be applied to other YSOs, pre-selected for hourly or faster near-infrared variability, at more wavelengths and more epochs. Such a program will measure the sizes of the inner disk walls and probe their time variability and azimuthal uniformity.

\section*{Acknowledgements}
The authors thank Rachel Akeson, John Carpenter, Kevin Flaherty, Lynne Hillenbrand, Heather Knutson, Patrick Ogle, Inseok Song, Karl Stapelfeldt, and Barbara Whitney for their contributions to the YSOVAR project and valuable discussions on this work. HYAM acknowledges the support from the IPAC Visiting Graduate Research Fellowship program. PP acknowledges the work of Nancy Silbermann, William Mahoney, and the Spitzer scheduling team in scheduling the data volume-intensive observations, and thanks Greg Doppmann for providing the Keck/NIRSPEC spectrum. KRC acknowledges support provided by the NSF through grant AST-1449476. AMW thanks the staff of the Observatorio Astron\'omico Nacional in Sierra San Pedro M\'{a}rtir. This work is based in part on observations made with the Spitzer Space Telescope, which is operated by the Jet Propulsion Laboratory, California Institute of Technology under a contract with NASA. Support for this work was provided by NASA through an award issued by JPL/Caltech.

{\it Facilities:} \facility{CTIO:1.3m, Mayall, OANSPM:HJT, SOAR, Spitzer (IRAC)}

\clearpage

\clearpage

\clearpage

\begin{figure}
\epsscale{.6}
\plotone{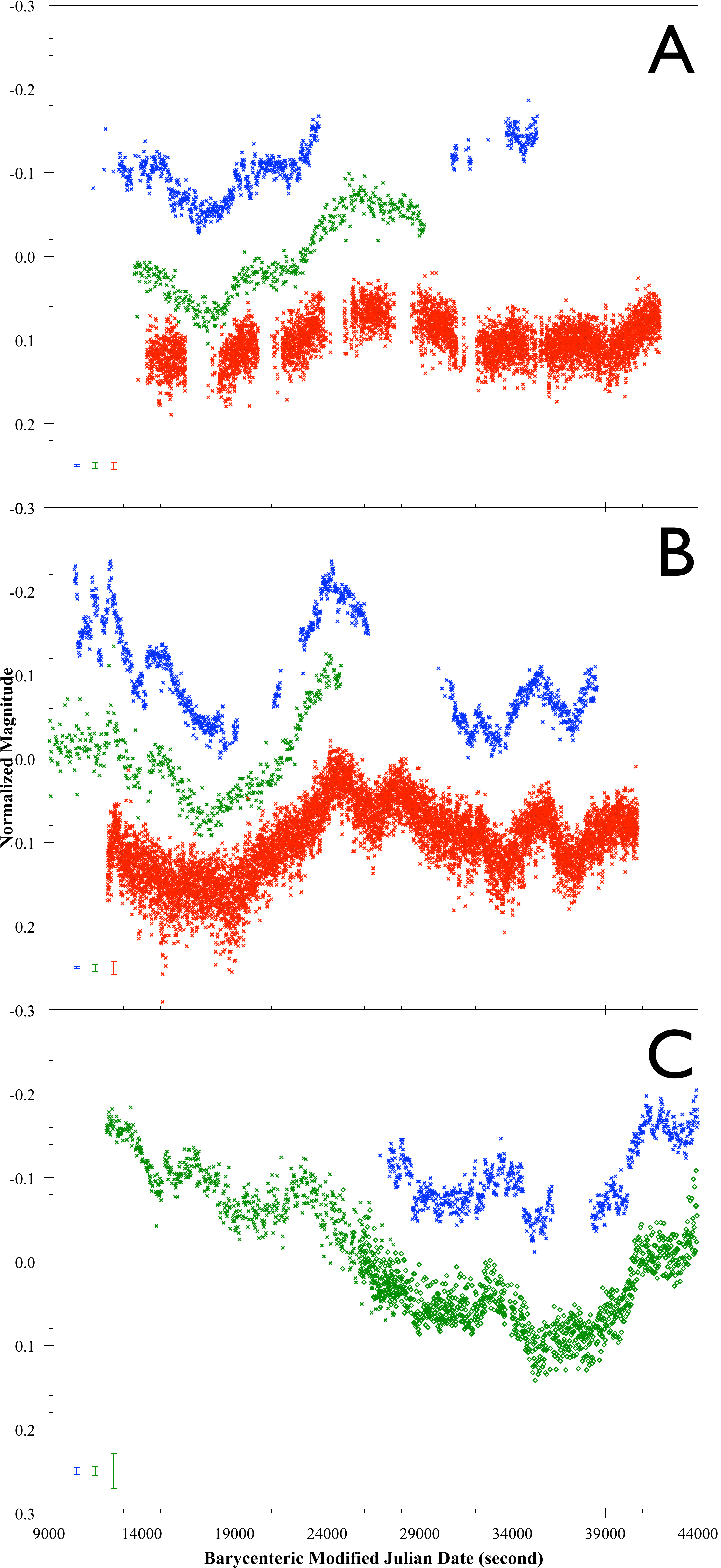}
\clearpage
\end{figure}
\begin{figure}
\caption{Normalized light curves of YLW 16B. Only the data used for the time lag measurements are plotted. The blue, green, and red points represent the $H$, $K$ and 4.5 $\micron$ data, and are shifted by $-0.1$, $0$, and $+0.1$ magnitude for clarity, respectively. The averages of the nominal photometric errors of each data set are shown in the lower left corner. (A) First night. The amplitude of the 4.5 $\micron$ light curve is too large for eye comparison and is reduced by a factor of 0.5 for illustration purposes only. (B) Second night. For illustration purposes only, the amplitude of the 4.5 $\micron$ light curve is reduced by a factor of 0.3. (C) Third night. The $K$ band data from SMART 1.3-m/ANDICAM and from OAN Johnson 1.5-m/Camila are plotted as crosses and diamonds, respectively.\label{fig3}}
\end{figure}

\clearpage

\begin{figure}
\epsscale{1}
\plotone{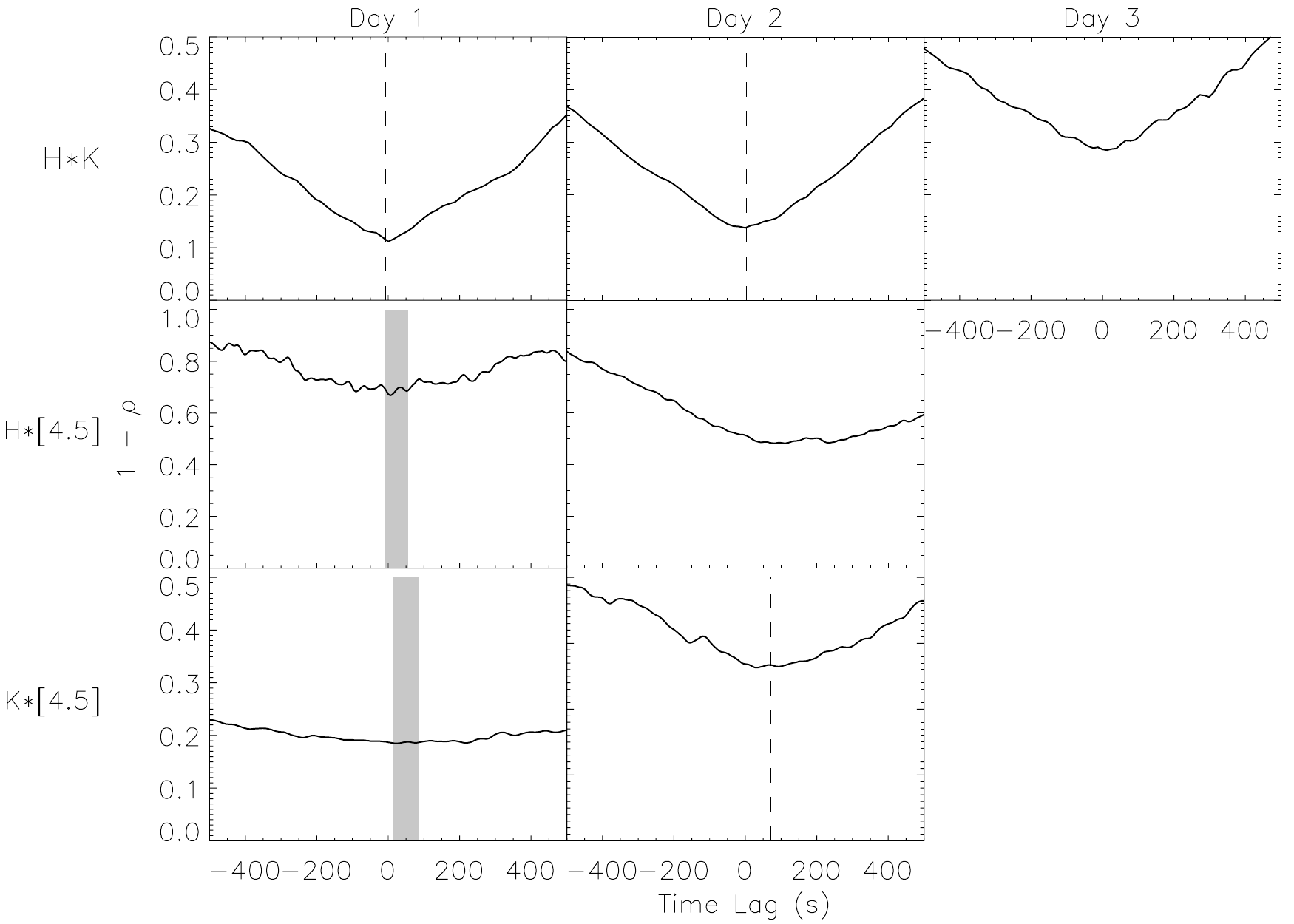}
\caption{Cross-correlation functions of YLW 16B between the $H$, $K$, and 4.5 $\micron$ time series. The data are shown in time lag (in seconds) vs $1-\rho$, where $\rho$ is the corresponding cross correlation function. the plots are aligned vertically for the day of observations, and horizontally for the pair of compared wavebands. The measurements of time lags with errors smaller than 5 seconds are labeled with vertical dashed lines; the 1$\sigma$ error ranges of those with larger errors are shaded. The $H \star K$ cross correlations are consistent with a zero time lag within the errors, while the 4.5 $\micron$ data lag behind both $H$ and $K$ by a consistent amount of time,  $74.5\pm3.2$ seconds.\label{fig4}}
\end{figure}

\clearpage

\begin{figure}
\epsscale{1}
\plotone{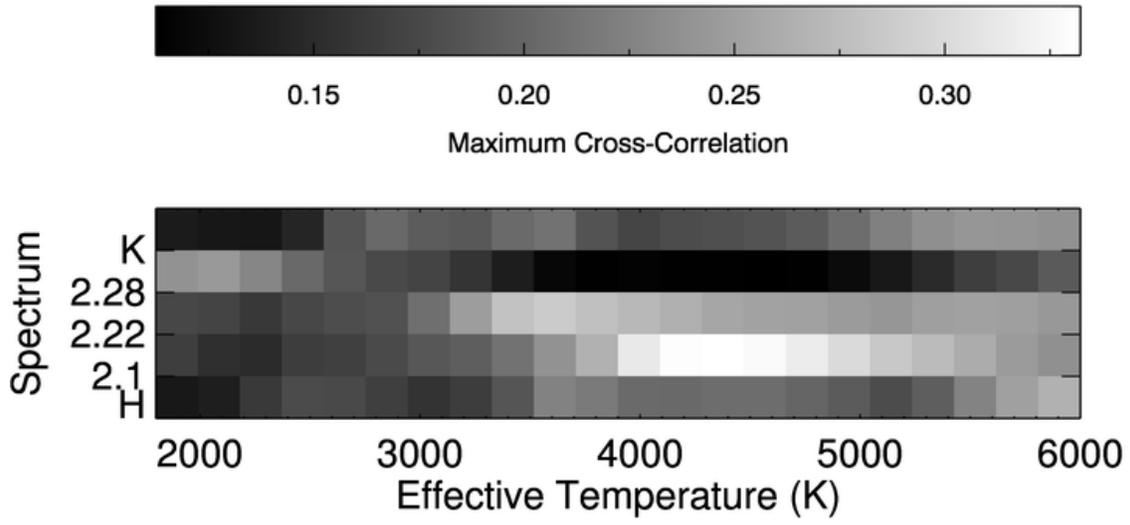}
\caption{Cross-correlations of the YLW 16B spectra with the BT-Settl models. From top to bottom are, sorted in descending order of wavelength, the combined IRTF/SpeX spectrum in $K$ band, three orders of the Keck/NIRSPEC spectrum near 2.28, 2.22, and 2.10 $\micron$, and the IRTF/SpeX spectrum in $H$ band.\label{fig5}}
\end{figure}

\clearpage

\begin{figure}
\epsscale{.5}
\plotone{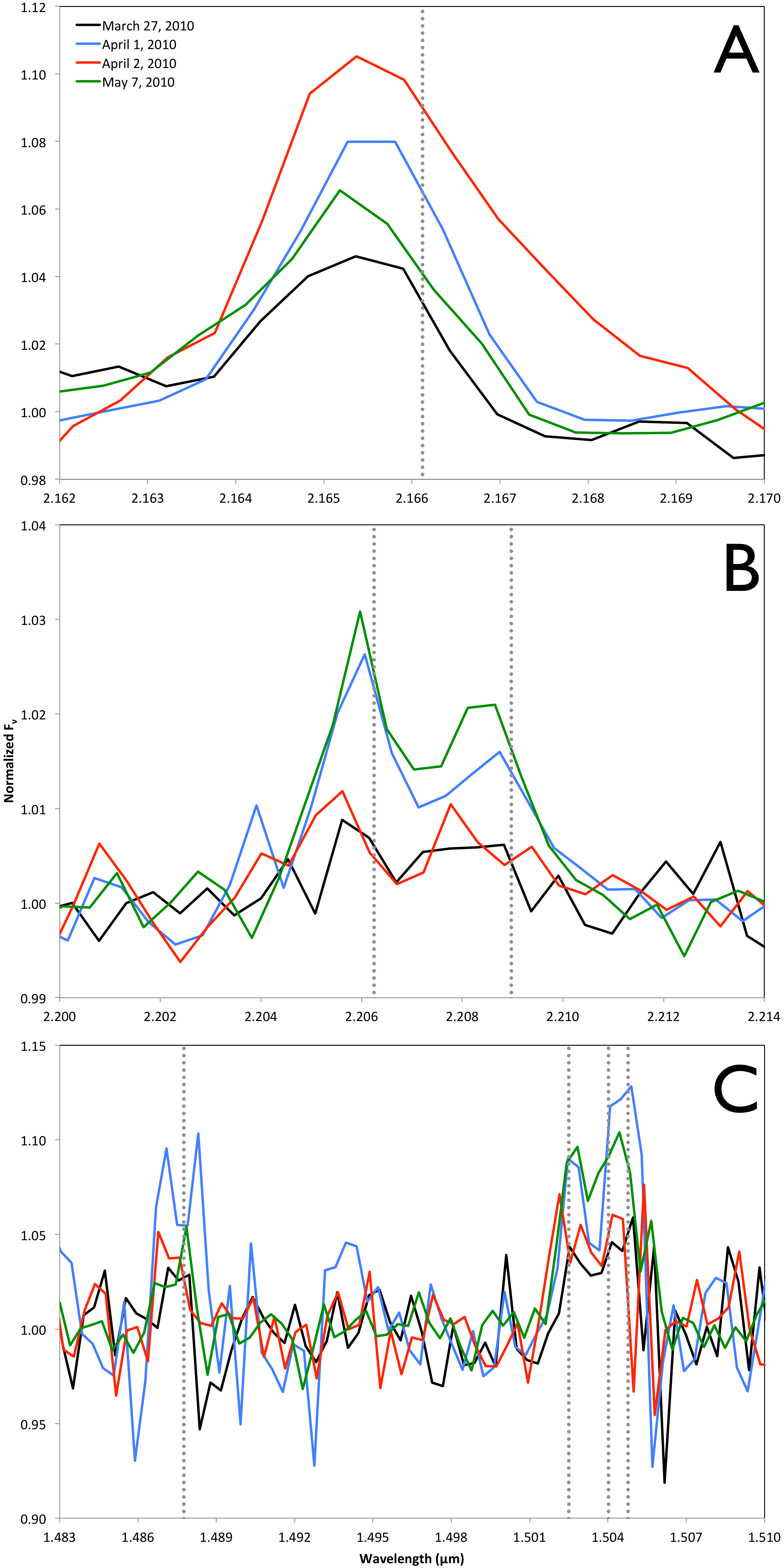}
\caption{Variations of the emission lines in the quasi-contemporary IRTF/SpeX spectra. (A) Br $\gamma$. (B) \ion{Na}{1}. (C) \ion{Mg}{1}.\label{fig6}}
\end{figure}

\clearpage

\begin{figure}
\epsscale{1}
\plotone{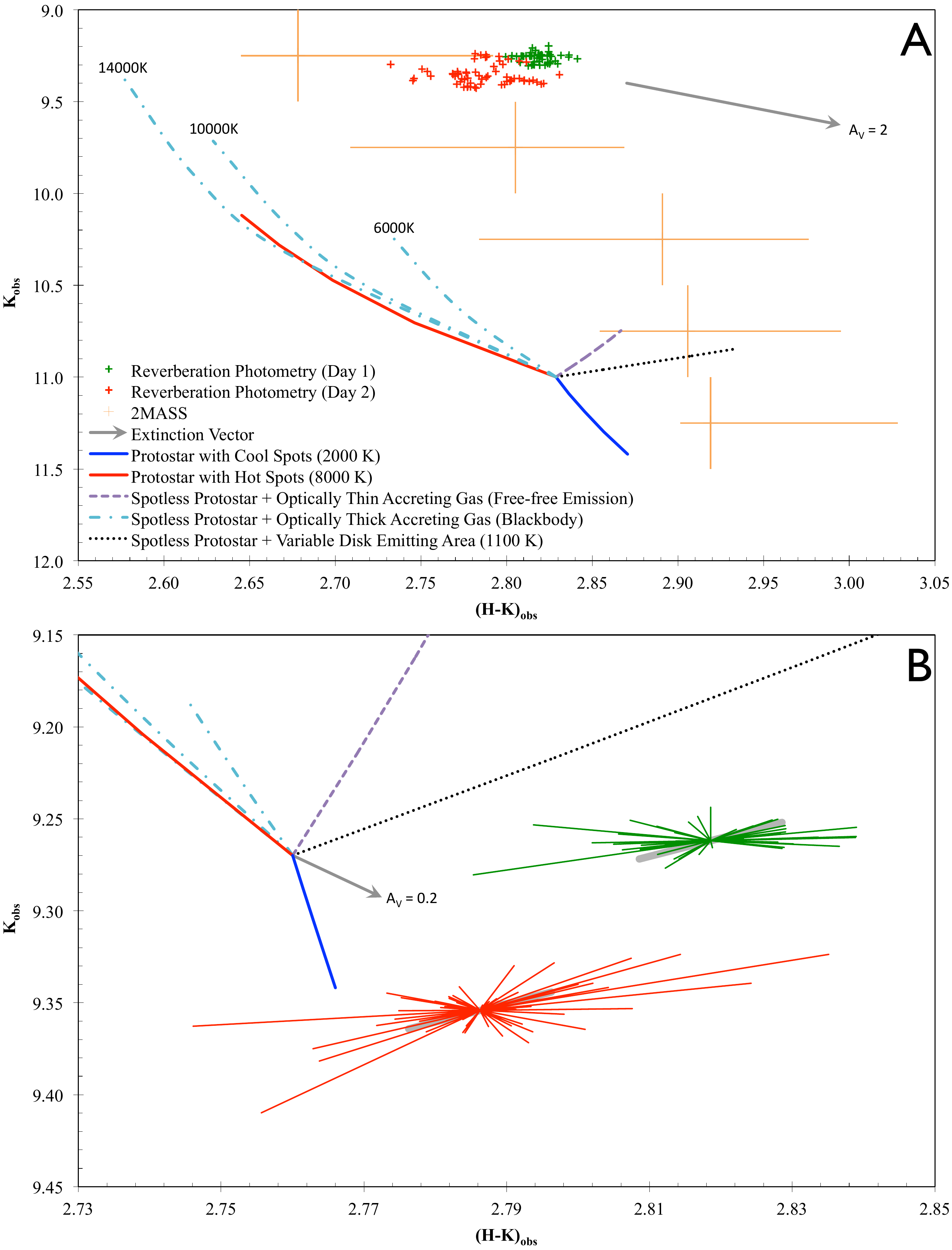}
\end{figure}
\clearpage
\begin{figure}
\caption{Color-magnitude diagram of YLW 16B. (A) Full data set. For the multi-epoch 2MASS calibration data, the vertical and horizontal bars show the bin size in $K_s$ and the first and third quartiles in $H-K_s$, respectively. They intersect at the mid-range of the $K_s$ bin and the median $H-K_s$ color. The green and red crosses represent the photometry in the reverberation experiment from the first two nights, combined with individual measurements every 204 seconds. This photometry is not calibrated to the 2MASS photometric system, and is subject to some shift in $H-K$ (and in $K$ to a lesser extent) when compared with the 2MASS data. However, the first two nights' data were obtained with the same sets of instruments, and are thus internally comparable. $H$ and $K$ photometry from the third night was made with different instrument sets and is not shown. The extinction vector is in grey. The variation models, as indicated in the legend, are discussed in detail in the text. (B) Zoom-in view of the reverberation photometry. The green and red polar diagrams show the amplitude and direction of the color-magnitude variation of each two consecutive 204 second-combined measurements, with their origins at the mean position of each night. The thick grey bar underneath each polar diagram illustrates the error correlation vector for $\pm$0.01 magnitude. This is 5 times larger than the nominal photometric errors in $K$, and should be the largest correlated error possible.\label{fig7}}
\end{figure}

\clearpage

\begin{figure}
\epsscale{1}
\plotone{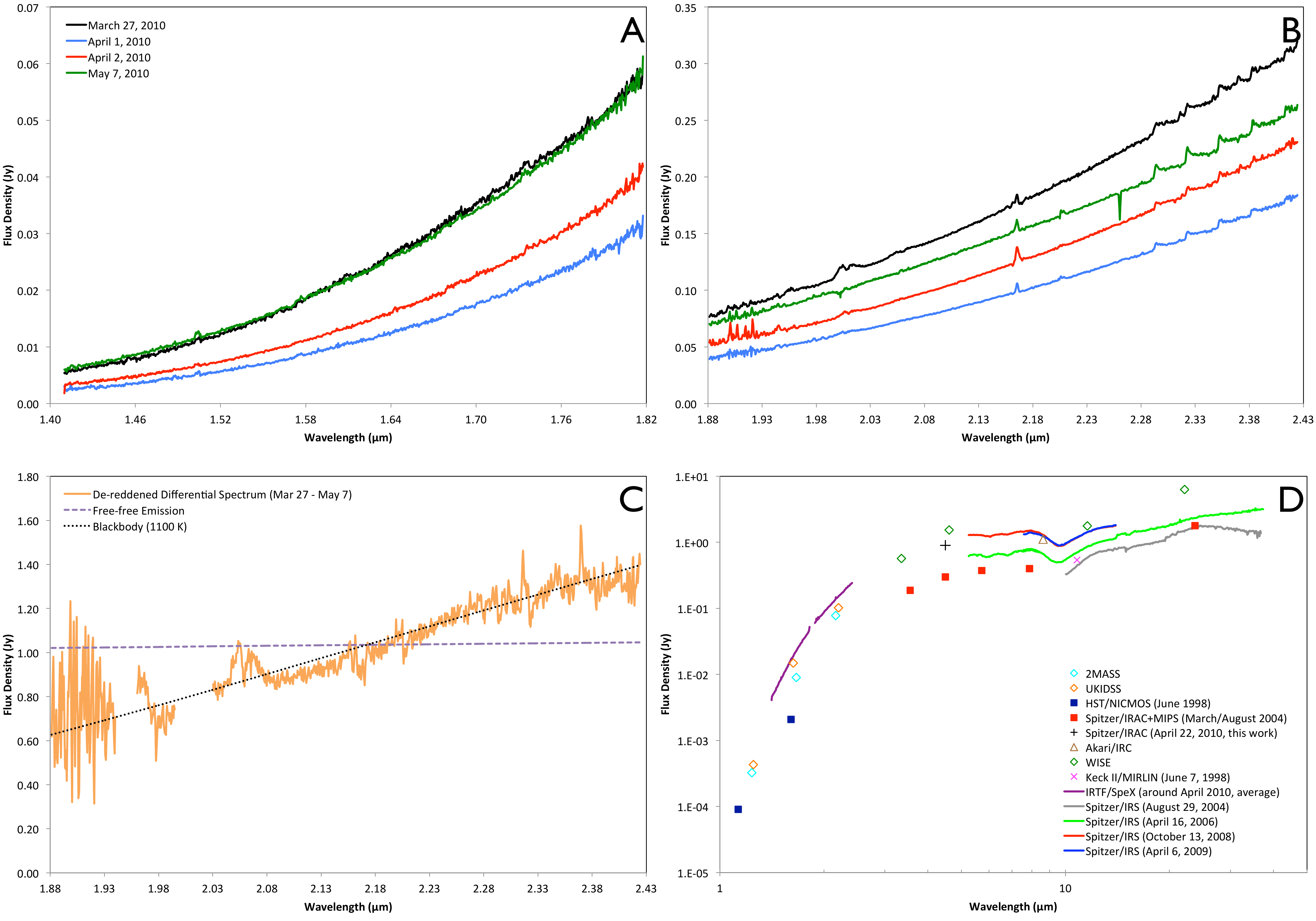}
\caption{Time variations of the quasi-contemporary IRTF/SpeX spectra. (A) $H$ band. The continua on March 27 and May 7 are nearly identical. (B) $K$ band. The continuum on March 27 is significantly brighter than on May 7. (C) Differential spectrum in $K$ band between March 27 and May 7, 2010, de-reddened and compared with the best-fit free-free and blackbody spectra. (D) Placing the spectra into the context of the near- to mid-infrared SED of YLW 16B, of which the variability over the years are evident. The 2MASS, UKIDSS (Data Release 9), {\it HST}/NICMOS, {\it Spitzer}/IRAC+MIPS, {\it Akari}/IRC, {\it WISE} (AllWISE Data Release), and Keck II/MERLIN data are based on \citet{skr06,law07,all02,gut09,ish10,wri10,bar05}, respectively.\label{fig8}}
\end{figure}

\clearpage

\begin{figure}
\epsscale{1}
\plotone{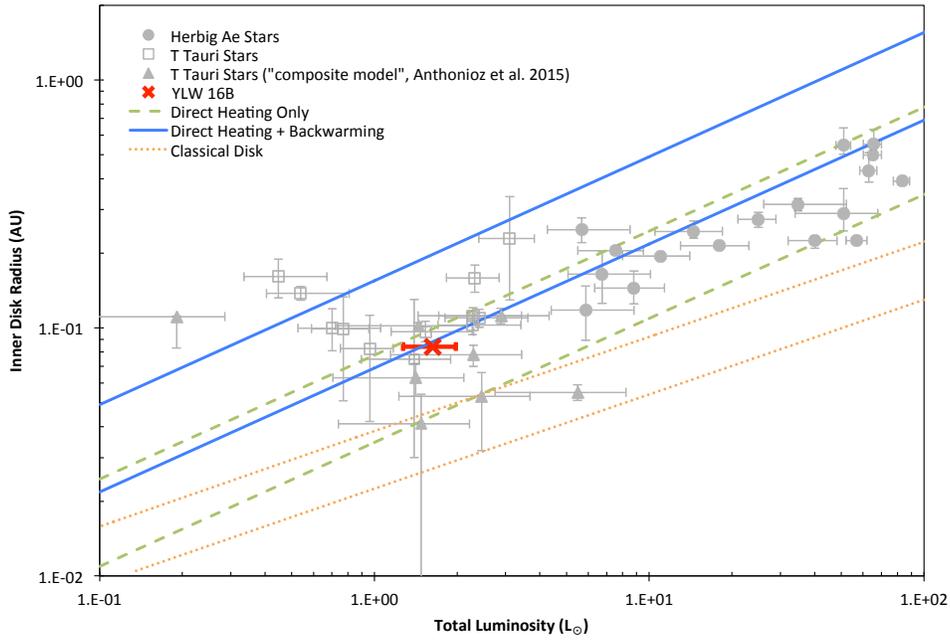}
\caption{Inner disk size-luminosity diagram. Comparison of the photo-reverberation measurement of the inner disk size of YLW 16B (red) with those constrained by near-infrared interferometric observations (grey, with different symbol styles for stellar types). The nominal uncertainty of the photo-reverberation measurement, 0.004 AU, is smaller than the symbol size and is not plotted. The interferometric measurements were adopted from \citet{mil07} and \citet{eis07}, combined with those in \citet{ant15} based on the ``composite model'' for both thermal and scattered emission if the data are fitted with a reduced $\chi^2 \leq 3$. For each sublimation model, the upper and lower lines represent the dust sublimation temperatures of 1000 K and 1500 K, respectively.\label{fig9}}
\end{figure}

\clearpage

\begin{deluxetable}{cccc}
\tabletypesize{\scriptsize}
\tablecaption{Index of the sources monitored in the reverberation experiment\label{index}}
\tablehead{
\colhead{Running No.} & \colhead{R.A. (J2000.0)} & \colhead{Dec. (J2000.0)} & \colhead{Other Identifiers}
}
\startdata
1 & $246.831300$ & $-24.694553$ & YLW 13A = IRS 40 = ISO-Oph 130 = GY 250 \\
2 & $246.839453$ & $-24.695297$ & YLW 13B = IRS 42 = ISO-Oph 132 = GY 252 \\
3 & $246.851625$ & $-24.696545$ & ISO-Oph 136 = GY 258 \\
4 & $246.852555$ & $-24.684280$ & ISO-Oph 137 \\
5 & $246.859523$ & $-24.712814$ & ISO-Oph 139 = GY 260 \\
6 & $246.860373$ & $-24.656410$ & YLW 16C = ISO-Oph 140 = GY 262 \\
7 & $246.860907$ & $-24.679209$ & GY 263 \\
8 & $246.862235$ & $-24.680784$ & YLW 15A = IRS 43 = ISO-Oph 141 = GY 265 \\
9 & $246.866781$ & $-24.659307$ & YLW 16A = IRS 44 = ISO-Oph 143 = GY 269 \\
10 & $246.872647$ & $-24.654497$ & YLW 16B = IRS 46 = ISO-Oph 145 = GY 274 \\
11 & $246.881403$ & $-24.640238$ & 2MASS J16273153-2438248 \\
12 & $246.887968$ & $-24.687576$ & YLW 18 = ISO-Oph 155 = GY 292 \\
13 & $246.888255$ & $-24.676695$ & 2MASS J16273318-2440361 \\
14 & $246.896922$ & $-24.642612$ & ISO-Oph 156 = GY 295 \\
15 & $246.897239$ & $-24.666201$ & 2MASS J16273533-2439583 \\
16 & $246.903374$ & $-24.660892$ & GY 298 \\
17 & $246.904062$ & $-24.700497$ & 2MASS J16273697-2442017 \\
18 & $246.905206$ & $-24.710560$ & ISO-Oph 161 = GY 301 \\
19 & $246.907594$ & $-24.646027$ & 2MASS J16273782-2438456 \\
20 & $246.910962$ & $-24.644220$ & ISO-Oph 164 = GY 310 \\
21 & $246.912269$ & $-24.672409$ & ISO-Oph 165 = GY 312 \\
22 & $246.914291$ & $-24.654318$ & ISO-Oph 166 = GY 314 \\
23 & $246.914796$ & $-24.725475$ & 2MASS J16273955-2443317 \\
24 & $246.915951$ & $-24.720852$ & YLW 45 = IRS 51 = ISO-Oph 167 = GY 315 \\
25 & $246.917064$ & $-24.643461$ & GY 317 \\
26 & $246.923345$ & $-24.643972$ & 2MASS J16274160-2438382 \\
27 & $246.927932$ & $-24.647396$ & ISO-Oph 172 = GY 326 \\
\enddata
\end{deluxetable}

\clearpage

\begin{deluxetable}{ccc}
\tabletypesize{\scriptsize}
\tablecaption{Differential photometry of all sources monitored in the reverberation experiment\label{database}}
\tablehead{
\colhead{BMJD (Second)\tablenotemark{a}} & \colhead{Magnitude} &
\colhead{Magnitude Error}
}
\startdata
\multicolumn{3}{c}{A1\_4\tablenotemark{b}} \\
\hline
13630.463 & 4.297 & 0.160 \\
13671.875 & 3.879 & 0.107 \\
13712.933 & 3.707 & 0.096 \\
13773.759 & 5.389 & 0.427 \\
13814.926 & 3.812 & 0.099 \\
13855.654 & 4.888 & 0.266 \\
13916.368 & 3.824 & 0.103 \\
13957.328 & 4.280 & 0.154 \\
13997.946 & 4.087 & 0.136 \\
14154.597 & 4.732 & 0.226 \\
14195.694 & 3.778 & 0.100 \\
14236.787 & 5.062 & 0.328 \\
14297.641 & 3.970 & 0.119 \\
14338.848 & 3.588 & 0.082 \\
14379.502 & 4.575 & 0.212 \\
14440.296 & 4.706 & 0.228 \\
14481.684 & 4.062 & 0.125 \\
14522.168 & 4.169 & 0.147 \\
\vdots & \vdots & \vdots \\
\enddata
\tablenotetext{a}{Mid-exposure time in barycentric modified Julian date in dynamical time (BMJD\_TDB) second of the day.}
\tablenotetext{b}{The file name format is ``instrumental initial + day number + underscore  + running number of the star'', where the instrumental initials are A = ANDICAM, C = Camila, F = FLAMINGOS, S = Spartan, and the running numbers are from Table~\ref{index}. For example, the file ``A1\_4'' is the ANDICAM light curve on the first night for star 4 (ISO-Oph 137). A file name that does not exist (e.g., ``A1\_1'') means that the star was not covered by the instrument on that night.}
\end{deluxetable}

\clearpage

\end{document}